# City NPC: Plan for Building a Million Person Martian City-State

**Vincenzo Donofrio**
Michigan State University, East Lansing, MI
Email: donofr19@msu.edu

**Meghan Kirk**
Michigan State University, East Lansing, MI
Email: kirkmegh@msu.edu

## Abstract

The task of designing a city-state of one million residents on the planet Mars seemingly approaches the limit of perceived human ingenuity. A city-state on Mars of this capacity requires us to conform to and master the simplest aspects of survival yet also allows the opportunity to handle depths that humans on Earth have yet to imagine. Because the city-state will not be the initial structure for inhabitants of Mars, a simpler colony will be instead, it is important to note the progression from the finalized colony to the city. This paper will briefly detail aspects of the successful colony, Colony NPC, to give credence to the ignition of the city-state, City NPC. Consideration of technical, economical, and societal elements of the city are also discussed.

## Location

Careful consideration has gone into choosing the site location as detailed site selection for human colonization is a concept entirely unique to Mars. Decades of research has focused on suitable locations for a potential landing spot because it is arguably the most important decision leading to success for the Martian society. The site needs to consist of low altitude and therefore accessible landing zones for Starships, an abundance of shallow ice deposits, tunneling/lava tube friendly rock quality, accessibility to critical resources, and preferably somewhat near to appealing landmarks.

Colony NPC settled near an expanded crater in the Arcadia Planitia region. This region has been found to possess a combination of accessible lava tubes, secondary craters that indicate readily available water ice deposits, and extremely flat land.[1] Additionally, it neighbors two of Mars' volcanoes including the largest planetary mountain in the solar system, Olympic Mons, allowing for a potentially very successful and accessible tourist attraction.

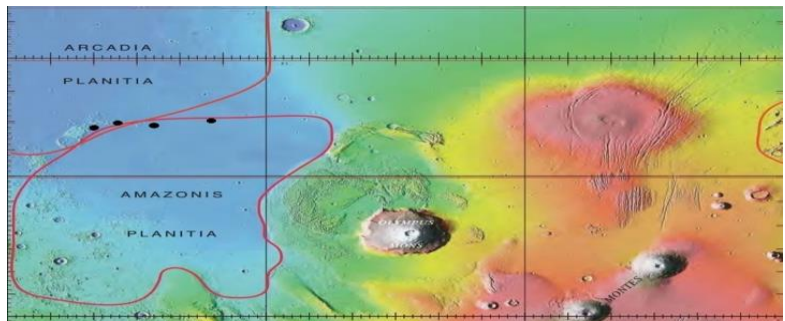
Figure 1: Northwest region indicating site location of Arcadia Planitia.[2]

City NPC will eventually expand its reach to the Martian moon Deimos. The smaller and outermost moon's main purpose will be to serve as the launch site for asteroid collection mainly due to its extremely low gravity and potential to produce in-situ fuel. Considering Deimos' size and makeup, a rudimentary colony will have to be sufficient enough to shelter the 100 or so specialists required in the asteroid mining industry; with majority of this paper detailing aspects of the city, Deimos will only be touched on again extensively in the *Asteroid Mining* section under *Economic*.

## Transition

With the focus now towards a Martian city, great excitement can be had in the presumptive success of the implementation of Colony NPC and prospects are positively favored regarding Mars, but it is important to keep steady and not rush the implementation of City NPC. It can be earnestly assumed that the colonists have mastered things such as resource harvesting, manufacturing, and ceramics; however, disciplines including in-situ crop creation and sustainable production may require another decade or so to refine. During this time, extensive greenhouses will become finalized and city outlining will continue to progress. Upon completion of this period, the city will have acquired the necessary infrastructure, power supply, and resource to begin accepting initial guests.

## Progression of Population Size

Before outlining the various technical aspects on the city, it is necessary to overview the general progression of City NPC to preface certain decisions made related to the pace of construction and importation. Colony NPC required fifteen years (2030-2045) upon conception to achieve a population of 1,000, with the maximum number of passengers per passenger Starships reaching 20. After the necessary five or ten years (2045-2055) detailed in the *Transition* section used to minimally prepare for intake, City NPC will accept incoming inhabitants. Efficiently following the Hohmann Transfer, cargo and passenger ships will land approximately every two years. Over the next decade (~5 transfers; 2055-2065), a total of 80 passengers will occupy each of the 10 passenger Starships per transfer. This means that the ships will only transport at 80% their maximum capacity as well as half of the maximum fleet size, allowing for further preparations yet enough of an influx in population size to measure quality of so-far constructed city. The parameters will then be set to 100 passengers per a total of 20 Starships, an intake of 10,000 people per decade requiring approximately 90 years (2065-2155) until one million inhabit City NPC. While it may seem like the rate of change in population intake reaches maxium linearity relatively quickly, City NPC is confident that the 35 or so years until linearity will be sufficient to match this progression. Additionally, the projected fleet size used here is heavily conservative with suggested reports, keeping in mind realization of current Starship technology. Of course, strictly cargo ships will intervalley arrive every two years upon conception of the colony.

## Overview of City

Prior to the first manned mission to Mars, many robotic missions will have taken place which allows the fortunate ability to specifically map out a city of a million before even the first thousand arrive. There will be five expanded craters located with sizes constrained to typically 5 kilometers in diameter. To account for the potential million inhabitants, between three to five 25 kilometer long and 100 meter wide lava tubes will need to be confirmed to closely surround each crater, and while seemingly a daunting task, the aforementioned selected region provides us with an abundance of lava and caves to choose from, with majority meeting the required dimensions. Tunnels of this size will allow around half of the future million inhabitants volume for housing, protection from radiation and meteorites, and temperature stability with only prior smoothing, cement sealing, and quality-assuring necessary with large-scale habitable volume digging unnecessary.

Colony NPC was constructed in only one tunnel using simply 3 kilometers of its length which housed all eventual one thousand colonists. City NPC will make use of all 20 expected habitable tunnels as well as all five craters to house the one-million-person city: each tunnel will contain 25,000 people and each crater will 100,000 people. Each tunnel will possess one medium-sized small modular reactor to supply power for the inhabitants, with each reactor accompanied by a large-sized lake. The finalized crater will be covered by an additive constructed geodesic roof and will consist of two nuclear reactors constructed above ground. Additionally, industrial and recreational areas will both inhabit the tunnels and craters.

The crater with the initial colony site located in one of its surrounding tunnels will serve as the capital of the city, hosting central executive offices. However, each crater will be regarded as a sub city, with the collection making up the entirety of the Martian city-state.

# Technical

**Habitat**

Lava tubes within Mars are similarly formed to those within Earth and the Moon: flowing lava recedes and leaves a lengthy tunnel many kilometers long in the rock it inhabits. Tunnel diameters have been observed to reach as great as 249 meters, substantially larger than those found on Earth simply due to Mars' lower gravity. The habitat for Colony NPC was constructed within a smoothed lava tube. For 1,000 colonists, the design was kept simple: a lave tube 100 meters in diameter and 3 kilometers in length.

City NPC will see its million inhabitants reside both in the expansive lava tubes underground and medium-sized craters that surround. Specifically, there will be five craters picked out in the Arcadia Planitia region, typically around five kilometers in diameter as stated previously. Considering the massive quantities of lava tubes formed in the region, it should not come with too much difficulty to find about four 25+ kilometer long and approximately 100 meters in diameter tunnels of required quality to inhabit surrounding each crater. Each crater will house around 100,000 citizens and with every tunnel filled to a capacity of 25,000. The city will look to take shape of a square with 100 kilometers a side to account for necessary spacing between the tunnels and craters.[3]

Using transported road headers and other excavation technology, the tunnels will eventually connect to the crater, aiding in atmospheric requirements for the dome as well further pushing along societal cohesion.

**Shelter**

Beyond the initial one thousand colonists, 3D printing will be used to create the residencies and workspaces. Large scale 3D printers will initially need to be transported, then can be made with in-situ created technology and metals. The Marsha structure created by the American advanced construction technology company, *AI Superfactory,* is the best choice currently for limited-sized habitats. This model can be slightly adjusted to contain five levels per structure (~25 meters in height), with each level possessing 34 $m^2$ of space. The current model (nearly 2.5 times smaller than this description) takes about 30 hours to complete which means City NPC's version will spend approximately three Earth days building each home. To allow for residencies to be available for incoming colonists, there will need to be 100 industrial-scale 3D printers imported after the finalization of Colony NPC, which will cost around $50 million to ship.[4]

The Marsha housing requires an innovative mixture of basalt fiber extracted from Martian rock and renewable bioplastic (polylactic acid) which will be processed from plants grown in greenhouses. Chemical analyses of rocks on Mars shows that the lava tubes and thus the Martian crust built upon them were classified to mostly compose of basalt, allowing for similar rudimentary extraction processes performed on Earth to be used for basalt production. A staple crop like maize will be a capable target to produce the necessary bioplastic; upon harvesting, the crop can be soaked and ground so the endosperm can be separated from the gluten and fiber which will be added with enzymes and then bacterial cultures to cause the produced sugar to ferment into lactic acid that can bond into the polymers needed for production.[5]

To make use of the depth of the crater, the choice of building to house inhabitants will be the skyscraper using again the 3D printed construction method. While not officially completed on Earth, there are plans to realize this method. An American large-scale 3D print company, Cazza, has plans to construct a skyscraper 40 or 50 stories high using concrete and steel in which the crane printing system can "easily be adopted with existing terrestrial cranes." This means that this process is essentially automating already constructed cranes to efficiently print large-scale skyscrapers. Residential skyscrapers of this magnitude can hold safely around 10,000 people, meaning a crater will need to create 10 of these buildings. Most materials can be made in-situ (as well as with the Marsha habitat), but two tower cranes will be necessary which, including counterweight, can all together weigh around 30,000 kilograms and therefore cost $15 million.

The craters will also consist of a geodesic dome made using glass panels and carbon fiber. To complete this, a modified crane tower sporting specification required for large dome building will be transported, estimating to be $50 million to match a quantity of around half of a dozen needed per completion.

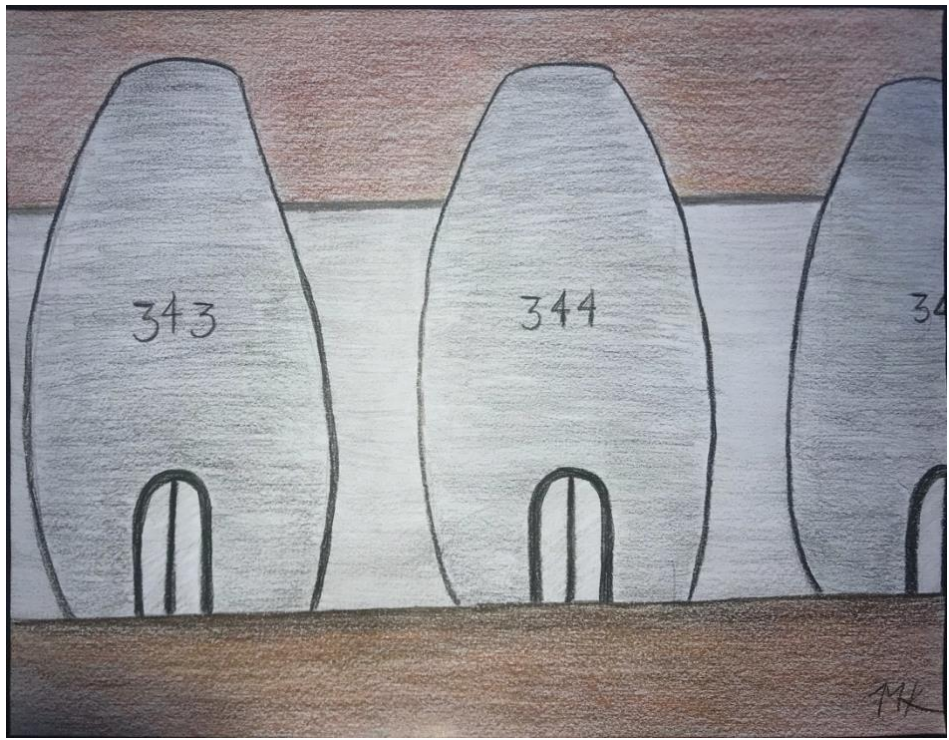
Figure 2: Marsha housing in a tunnel connected in condo-like array.

**Power**

Colony NPC required two kWe per member of the colony, which was mainly supplied using twenty 100 kWe NASA Kilopower reactors in addition to a minimal input of solar energy. Considering the cost and effort to maintain even general upkeep of the solar panels necessary for its power to be relied on any further, City NPC will not expand its use, only maintain the initially installed source which will eventually serve solely to upstart the reactors and other extremely trivial tasks. Instead, the city-state will transition entirely to nuclear power, specifically by way of nuclear fission - as it more compact and easier to deploy than a fusion reactor.

Judging the requirements needed for the colony, City NPC will need at least two GWe of power supplied daily. This supply will increase by a factor of 2.5, taking into account that the eventual one-million-person city will become slightly more energy intensive over time than the colony. The inflated requirement of 5 GWe is sensible considering that the minimum energy requirement for the colony is detailed in the *MarsSociety* reference at two kWe per member which only considers a limited amount of supply needed. However, we will slightly increase the per capita requirement to 5.46 kWe to fit the model of the main nuclear reactors.

City NPC will utilize the Integral Molten Salt Reactor (IMSR) developed by the Canadian company Terrestrial Energy. The IMSR is a safer and more efficient design than water-cooled reactors, with its already liquid fuel mitigating risk of meltdown and its operation at atmospheric pressure eliminating the possibility of explosion. Its core-unit is designed to be completely replaced after seven years. Cooling occurs in the molten salt loop, with a subordinate salt circuit propelling the steam engine by transferring heat to the water loop. The IMSR utilizes a thermal-spectrum, molten-fluoride-salt reactor system with low-enriched, standard-assay uranium as fuel – with this quality of resource mathematically predicted to be plentiful on Mars[6].

There will need to be forty 141 MWe, the second largest model available from T.E., IMSRs transported to Mars to power the city: one in each tunnel (20) and two to power each crater (20). The tunnel-bound reactors will be constructed at the midway point (approximately 12.5 kilometers in) to meet the simplistic climate and temperature control requirements, while those used for the craters will be planted above ground, but adjacent to the crater to avoid difficult and exhausting control parameters. While the total transport mass parameters are not currently known for the IMSR, we can assume it will occupy the size of a typical medium-sized MSR of approximately 2500 metric tons and the total cost will be $50 billion; additionally, it will be safe to bring subsequent 5 MWe basic backup reactors in case of mass failure occurring in any section of the city meaning another $1 billion will be accounted.

**Atmosphere and Climate Control**

The lava tubes' will be occupied first and earlier on in the progression of the city and thus their atmospheric composition will not be as fine compared to the craters. An example to follow for the tunnels is to consider the composition used in NASA's initial space station, Skylab, which was composed of 74% oxygen and 26% Nitrogen. Important to note here is the contribution nitrogen will have on the plant life in the tunnels' greenhouses. A symbolic notion as well as steady progress towards terraformation is to implement an atmosphere that begins to resemble that of Earth's. Each crater will be composed of 78% nitrogen, 21% oxygen, and 1% carbon dioxide. While nitrogen and oxygen will serve as the most productively used gases, an influx of carbon dioxide and similarly resembling an Earthling atmosphere represents a step forward to the ultimate Martian goal.

All gases can be produced in-situ through various processes. Oxygen is produced through water electrolysis where approximately 700,000 kilograms will be consumed for a population of one million, with roughly 790,000 kilograms of water necessary for full replenishment. Nitrogen rich minerals on Mars are thought to accessible in top layers of the soil of mainly the western hemisphere were relatively volcanically active regions reside. As inhabitants of this territory, the citizens will be able to locate with some ease the silicate minerals that lock substantial amounts of environmental nitrogen within magma or lava, as well as the chondrites that are able to contain roughly 28% of nitrogen per weight. The carbon dioxide present within the crater can initially be filled by passing through the carbon dioxide created in the tunnels by the colonists through airlock devices, with the remaining released from the tunnels to the atmosphere through pipes to prevent imbalances and health risks.

Considering the chosen primary nuclear reactor design produces 300 MWt and 141 MWe, controlling the climate within the tunnels will come with some difficulty. This difficulty can be mitigated by correctly positioning the reactor along a point in the tunnel to allow for a quality heat cycle to generate. In each tunnel, the reactor will be placed at the midway point near the large lake. If the lake is at a relatively cool temperature, a proper ventilation system can be installed to propel air out to both sides unto the stream reaches the labor sections upon either ends of the tube which will then propel warmer air back to the lake. Even with a very wide and deep lake, an influx of 441 MW of total energy will give rise to its water alarmingly, which means that the city will need to make use of the abundance of water ice deposits in the Arcadia Planitia region to daily lower temperature.

**Food**

As previously mentioned, upon initialize colonization, there will be approximately 25 years' time needed to study, experiment, and implement correct measures to become entirely self-reliant on Martian-grown food, not to mention the similar amount of time needed to construct even the minimal amount of greenhouse infrastructure required for the city. So, the one thousand Martians occupying the planet during this period will require a total of 175,000 kilograms per year of typical dehydrated food imported at a total cost of roughly $2.5 billion.

The Martian diet will be dominated by a plethora of terrestrial staple and non-staple crops which will be directly suited for the eventual iron-rich soil derived from cleansed Martian regolith and necessary fertilizer. A several year experiment performed by Wageningen University and Research in the Netherlands used a "NASA-developed reproduction of Martian regolith" mixed with leftover compost from previous harvests to produce successful returns specifically from tomato (claimed as the "first tomatoes ever grown on Mars soil simulate turning red"), radish, rye, quinoa, and chives, and maize. Experiments like this and various others allow the relatively conservative assumption that with slight adjustments to fertilizer and regolith composition, most terrestrial crops can be grown in greenhouses on Mars.[7]

To provide all nutrients necessary for a healthy and sustainable diet, the inhabitants will need to import livestock. The simplest method would be to transport minimal amounts of female chickens, rabbits, and goats as well as maximal amounts of necessary frozen sperm to allow for population.

A general overview detailing the necessary requirements needed to feed the city of one million can show the promise that simplistic terrestrial crops possess. City NPC will require approximately 15,000 hectares to supply the whole population and with the aim of a relatively limited 1600 calories per day diet. While more crops will be harvested, preferably on seasonal basis, we can simply consider three choices to portray sustainability: quinoa (cal / kg: 1221; yield: 10 ton / ha), rye (cal / kg: 3380; yield: 4.5 ton / ha), and maize (cal / kg: 3840; yield: 6 ton / ha). Averaging between the three crops, a calorie supply of nearly 2500 calories per day per person will typically be available, enough to safely account for each individual.

The greenhouses to grow crops in can be constructed entirely from in-situ resources: concrete, steel, and glass. Soil can be created by cleaning the surface-layer regolith of any perchlorates and adding the correct amount of fertilizer for an intended crop. As previously stated, an advantage of approximately 25 years will be available to not only perfect greenhouse technology, but to periodically implement them at a rate exceeding that of the population.

**Metals and Various Material Production**

To minimalize import costs and progress to a self-sustainable city-state, City NPC will need to produce all manufacturing material in-situ requiring only initial transportation of machine equipment. The majority of resource collection will come by way of mining, requiring various electrical crawler-mounted roadheads to subtractive construct. Rock that will need to be excavated to lead from the tunnels to the craters with serve as initial mining sites, a positive notion considering the predictable rich mineral deposits near the impact positions. Upon completion of these sites, the mining team will continue with favorable zones either within or near the tunnels which will be scoped out during prior settlement missions. Roadhead technology and other necessary production equipment will cost around $1 billion to efficiently maneuver the city.

**Metals**

City NPC will utilize steelmaking to construct the Hyperloop railways as well as varied use in infrastructure. Earthling Steelmaking requires a Martian plentiful source of iron oxide but also a vacant Martian source of coal. To counteract this deficiency, hydrogen gas made from water electrolysis, a process used substantially in City NPC. The Swedish lead initiative, HYBRIT, endeavors to revolutionize steelmaking and enter production virtually adding no carbon footprint by using hydrogen gas instead of fossil fuels. Their goal is to "have a solution for fossil-free steel by 2035", allowing plenty of time to be prepared for use in City NPC.[8]

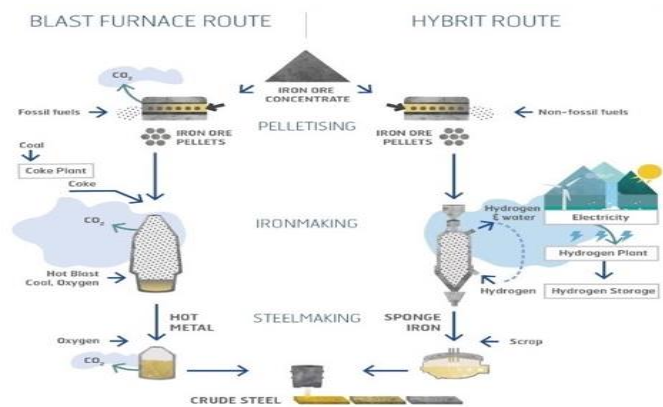

Figure 3: Detailing the process required to produce clean steelmaking.[8]

**Glass**

Glass production will primarily be implemented as windows in residential areas and greenhouses as well as panels used in geodesic dome coverings. About 40% of the weight of the soil on Mars is made up of silicon dioxide: the basic component in glassmaking. Unfortunately, Martian soil is also heavily made up of iron oxide as previously mentioned, which will leave the city with a lower quality of glass. However, this isn't really a problem for the citizens who will choose to opt for the most efficient processes, and the need to remove the iron oxide through a tough process isn't necessary. Therefore, the tinted silicon dioxide glass can be made using the same sand-melting techniques performed on Earth.

**Others**

Concrete can be used in some small-scale industrial infrastructure but will mainly be provided to the skyscrapers. Mixing heated, liquidized sulfur with Martian soil as an aggregate can create a concrete mix upon cooling; the preferred mix will be 50% sulfur and 50% soil at maximum size of 1 millimeter. Martian meteorites suggest elevated amounts of sulfur in their interior and Martian surface deposits are found to contain high levels of sulfur (averaging 6 wt% $SO_3$). Mixing iron oxide that will be mined for steelmaking can also be added to create a sturdier concrete composition. Inorganics like sand and clay can be produced with ease, citing basic techniques already used on Earth.

**Communication**

Communication during the human settlement of mars will be key. The routes that need to be covered are communication between Mars and Mars, and Mars and Earth and locations like the asteroid belt.

The most important form of communication, especially early on, will obviously be between Martians. Ideally for the few who are still left with exploration jobs, but really useful for everyone, will be an accessible way of communication that is at the push of button and attached to one's spacesuit. To allow the radio waves to be transmitted between Martians, satellites will need to be implemented, something not entirely too difficult, as long as there is the means of transportation to get the required position in Mars' low orbit. This is something that can be implemented during prior, unmanned missions, with the cost being virtually nothing when added to the total amount of the entirety of these missions to Mars. Communication between Martians and Earthlings and those at other locations can be done through these satellites as well, specifically using laser transmission techniques to limit the transmission distance handicap. Another option for this long-distance communication would be to use smartphones or the like to simply message between planets. This will of course require an internet connection to be founded on Mars, something that may actually impose greater difficulties than setting up satellites, but once invented will be used as another form of communication unfortunately again with a delay. Instantaneous communication between planets will perhaps be unearthed, but for now a slight delay will have to do.

**Transportation**

Transportation can be broken down to two aspects: an inter-tunnel and inter-crater system. The tunnels, separated into two distinct sections of approximately 12.5 kilometer in length, can be accessibly traversed using simple bicycles made entirely through sustainable materials and ridden upon concrete sidewalks. An inhabitant will need not spend longer than 30 minutes riding at a moderate pace to cover even the entire length of their section, with most needing only around 15 minutes to cover their required distance to a workplace or home. A typical inhabitant will be drawn to bicycling to also account for a large load of their physical activity completed while simultaneously traveling to their profession. Bikes will be the sole choice available within the relatively short end-to-end distanced craters.

A second option available within the tunneling shelter will be a version of the popular *Hyperloop* concept ever so ramping up in production. With current pod designs reaching speeds close to 400 km/hr, this mode of transportation will provide crucial, high-speed access to the entirety of the 25-kilometer lava tubes. Within the tunnels, it will be necessary to provide four pathways: two spanning the 12.5 kilometer sectional areas of the tube and appearing on the interior between the two rows of housing and two spanning the entire tunnel length located behind each row of housing and eventually passing around the large lake in the midway point. The reason to bifurcate the interior track is to define the two as solely 'inter-sectional' transportation; the remaining two lying on the exterior should provide the inhabitants with the necessary inter-tunnel travelling, however will primarily be used for inter-sectional movement, as well. These options, combined with the availability to bicycle, will provide an efficient flow of traffic and timely transportation complex.

Inter-crater transportation will additionally utilize and consist of hyperloop technology, perhaps subsequently allowing for its maximal use. Because of the typical 50 kilometers, above ground distance between craters, the only option will be high-speed transport. While it is important to consider outlets to aerial travel, such as singular-occupied drones, this technology on Earth is much further behind than even the progression of the hyperloop, allowing for it to be used again here. Tracks upwards of 100 kilometers can be installed on the Martian surface and allow for accessible movement between sub cities, where a journey of even this length would spend only a ridiculous 15 minutes of one's time.

It is important to note that two types of installation of the hyperloop design can be implemented sublayer and above ground. Considering the incredibly thin air pressure comprising above ground territory, air resistance will not be a factor. This allows pods on Martian surface to reach supersonic out in the open, with only an air bearing track beneath them. Because of the Earth-like air pressure required within the lava tubes, the loops there will need to add low air pressure tubes to allow for supersonic equivalent speeds. While presumably first being implemented some dozens of years after initial colonization to begin to accommodate for a drastic influx of inhabitants, it can be assumed that the air bearing track design and technology can be sustained using in-situ resources like various sources of pressurized gases as well as concrete or steel to build the track. However, it is likely the individual pods will need to be imported from Earth; a total of 88 pods will be required to accommodate final plans of the city, thus creating a total of $220 million in import costs. Finally, all cross or outer planetary distance (tourism, specific resource collection, Deimos missions) will be covered using specified Starships.

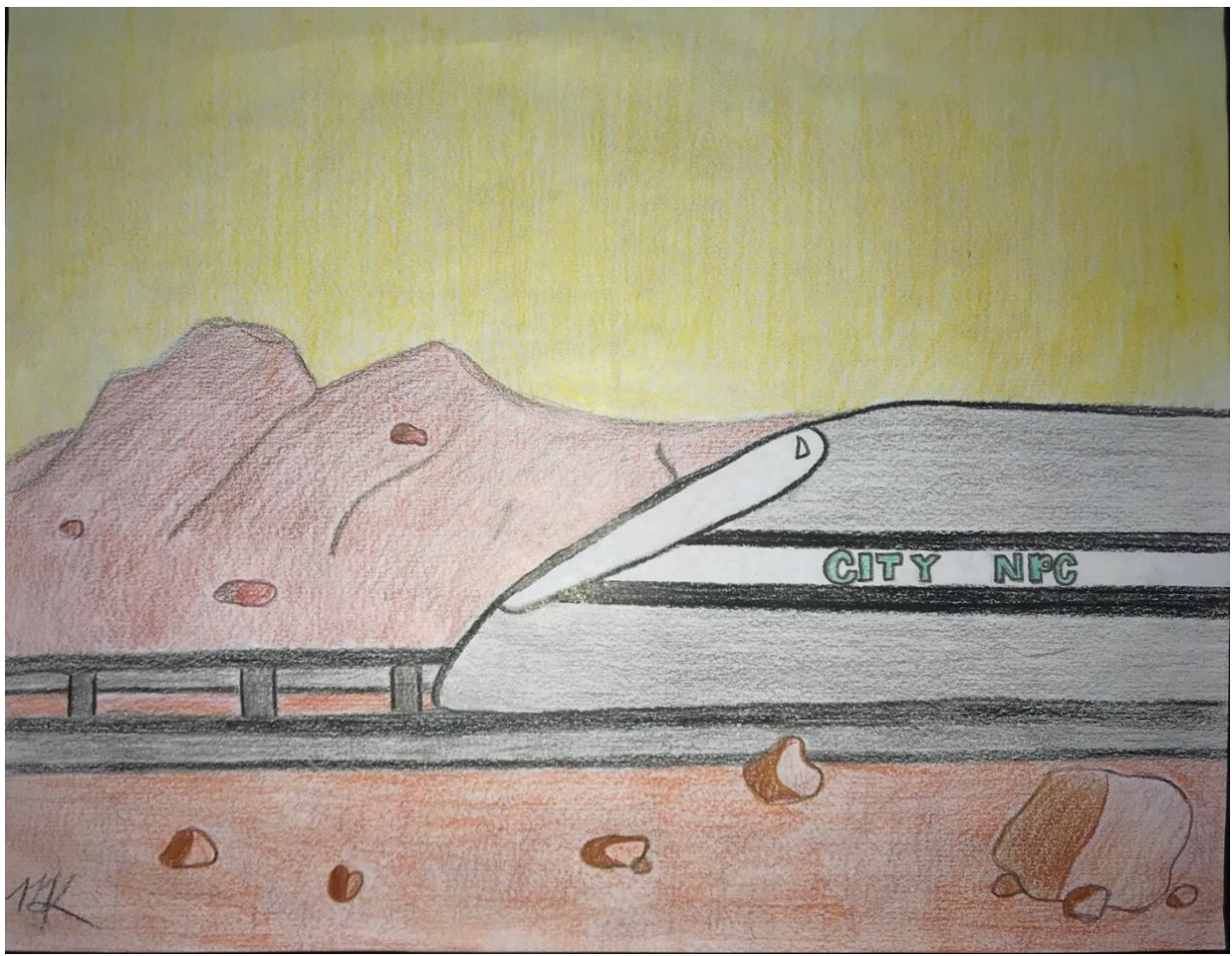

Figure 4: A City NPC Hyperloop zooming past.

# Political

The chosen government and political system will look very different between colony and city. For Colony NPC, it was expected to be the case that a private company (like SpaceX) with help from a leading nation and its space program (like the US and NASA) would lead the crew, implementing a local government on Mars comprised of a democratically selected leader and board of officials as well as a decently strong presence from said leading nation's government back on Earth. This plan made sense for a colony and can be related to the system deployed in colonial America: a composition of a few forms of local government, with also a great deal of interference from the British government (except Colony NPC did not face authoritarian rule from the Earthling government). These few colonists therefore possessed rights such as the ability to retain the rights to their creations: an important notion because of the expected boom of unique industries and opportunities for great inventions and achievements. The goal was to allow the chosen Earth government to oversee the colony, protect the rights and interests of its colonists, collect on the fairly priced export and import taxes, all with the confidence that Mars will quickly become economically, technically, and socially successful which will then lead to eventual self-governing metropolises. This initial high-level interest from Earth would always mean to decline steeply to coincide with the expansion of the site.

As City NPC approaches one million persons, it will develop into purely a participatory democracy. Majority of the citizens at this point on Mars will be expected to effectively contribute to the city through productive or unique labor, a characteristic completely unique to Mars. A strong central governmental presence will arise during this period on City NPC, acknowledging the necessity to break away from Earthling presence. Reaching this level of autonomy will import socially rewarding confidence into the city, indicating that the natural steps of progression are occurring.

Specific governmental guidelines are outlines below, detailing expectations from City NPC's political system.

**City NPC Governmental Guidelines**

1. **Executive:** A President will be elected by all eligible citizens by way of random ballot voting. To avoid complacency, yet to allow for a lengthy, successful presidency, presidential elections will occur every three years with the possibility to be twice re-elected. Candidates must be at least 25 years of age and meet basic civilian requirements: No convictions of any wrongdoings, good standing in community, et cetera. A President's focus will therefore shift from their previous professional regiment to now spending majority of their efforts leading the city-state (although engaging in past interests if vital to the city-state will be allowed with available time). The executive board will be appointed by the President with subsequent referendum occurring to vote out appointments meet with majority disdain. Board requirements will be exact to those necessary for presidential candidates.
2. **Local Executive:** Each sub city of 200,000 citizens will elect a mayor and subsequent council. The role of Martian mayor will be similar to the mayoral role found in many cities in the United States: preside at council meetings and act ceremoniously as the head of the city for community driven purposes; however, upon necessity, the mayor and council can regionally administer a vote for legislations spontaneously. While open to all citizens especially inspired to unite their sub city, mayoral candidates will be encouraged to reach elder stage life and pertain an extensive history of social justice engineering and facilitation. Elections will only commence if a bi-yearly referendum concludes that a majority favor the removal of the mayor and/or their council.
3. **Judicial:** The judiciary will solely consist of a national counsel based on the assumption that legal cases will be rare but supreme upon occurrence. A panel of 11 will be elected every six years, with unlimited re-election a possibility: a competent judge providing unambiguous ruling warrants stay. Candidates will be encouraged to spend time either as part of a local council or national board but will be available to all backgrounds. Regarding judicial hearings, citizens determined guilty will not be tolerated and will result in deportation to Earth and permanent ban from City NPC.

4. **Legislative:** All legislative markings will be facilitated by regional or national council and administer a local or universal vote from required citizens, respectively.

**Political Progression**

Considering the city will be partitioned into five sub cities of 200,000 people, it can be expected that perhaps all five areas will be shifted towards different points upon the standard political spectrum and, in fact, should be encouraged. Of course, even though each local government will strive for health, it can also be expected that some systems may fail. The strongest governments will inherently survive over time, a length of time however that may persist longer than expected. While the objective isn't to deem or seek cities or sub cities as utter failures, those communities that reach harmful radicalization (collapse into an authoritarian state) will need to be immediately rehabilitated. Regardless, great speculation, especially from Earth, will fixate on the progression City NPC retains politically, with idealistic hopes garnering for a perfect society.

## Economic

A city-state containing one million people undoubtedly offers various routes on path to a booming economy. However, in order to maximize any financial gain brought to the city and its investors, it is important to find the perfect balance between efficiency and profitably. This balance will likely differ from colony to city, citing a progression from novelty to sustainability. However, it is important to note the economic environment surrounding the colony, to not only allow eventual understanding of any differences, but to track the realistic profits setting up City NPC.

Colony NPC cost an initial investment of around $10 billion, which mainly comprised of basic food, shelter, maintenance, advanced electronics, and salary. To pay back and exceed this investment, the colony focused on exporting novelty and collector items like Martian sand and rocks ($5 billion) and some readily accessible and in-demand precious metals or resources ($500 million). The small colony also made a fortune off the rise of profitable inventions disinterred as a result of this newfound challenge and territory ($10 billion-plus). Collectively, Colony NPC garnered a true return of approximately $5 billion, which conveniently set up the explained transition period from colony to city.

As City NPC's population reaches one million, some of these same techniques will be used only on a much larger scale. New business, though, will arise including tourism and travel, media and advertising, expanded research, and asteroid collection and mining. It is first necessary to detail the collection of expenses paid to fund a secure city.

**Expense**

Expenses documented, as well as returns detailed later, will span from the initial acceptance of city inhabitants (2065) unto a population of one million persists (2155) because intrinsically Martian economic culture will require fine consideration and thought. While intriguing, it is important to not try and account for, vaguely and halfheartedly, economic potential beyond this span, but instead deterministically outline the necessary steps for success simply for this specified city-state, as majority of focus up to this point has resided in this mindset.

Majority of expenses will fall under three categories: salary, import, and investment. We will assume that for the first 10 or so years of the city's existence salary expenses will be covered by the state, which means $100 billion will need to cover the one hundred thousand inhabitants and their typical one hundred thousand-dollar yearly wages. Thereafter, we can expect the amount of state-issued salary to drop to an eventual 33% of the population, as the private sector progresses.

Importations will be prevalent especially during the first couple decades until they then decrease at a rate corresponding to the rate of self-sustainability. It is also wise to make note of investments made possible by supportive governments or corporations which will mainly supply the creation of the transported technology necessary for survival.

Imports costs over the first 100 years or so will include: $50 billion for the reactors, $100 million for industrial-sized and tower crane 3D printers, $220 million for Hyperloop pods, among various others. Inflating the total import expense to $100 billion should cover all transported goods necessary for City NPC. For salaries, approximately $2.5 trillion dollars will be state issued to cover some of the eventual million inhabitants of Mars. A total investment expense of roughly $25 billion is reasonable, citing respectable participation from multi-billion-dollar companies. Overall, total expense costs for a million-person city will fall around $2.5 trillion dollars.

While seemingly an egregiously large number, it is very important to consider that this projection is over a gradual period of 100 years' time. During this time, the self-sustainability of the city will exponentially increase, meaning that the city will quickly decrease its total import costs. Additionally, booms in the tourism and asteroid mining industries described below around 50 years or so years after initial implementation will secure the path to a financially stable city.

**Return**

**Tourism and Travel**

As the construction of the luxury skyscrapers reach completion and the city-state can truly house one million people, the tourism and travel industries will begin to thrive. With five total luxurious hotels available at ten thousand people per building, City NPC will be able to accept fifty thousand tourists and travelers at a time. A Martian city is undoubtedly one-of-a-kind and leisure-inspired guests will flock to the idea of a holiday on Mars. Guests can expect to experience not only a stay on the planet, but also extensive, Starship-led tours of multiple landmarks on Mars, especially those near City NPC's region of Arcadia Planitia. Nearest to City NPC is Olympus Mons, the solar system's largest volcano, and a little further beyond that is the second largest system of canyons in the solar system called Valles Marineris. Not to mention the belt of three large shield volcanoes in the Tharsis region in between the two previously mentioned sites appropriately collectively titled Tharsis Montes. Considering the eighteen months or so needed to complete a round-trip from Earth-to-Mars-to-Earth, guests may feel compelled to stay a convenient six months; however, most would feel it necessary to give Mars a few years to truly absorb all the unique get-away, Mars, offers. During this time, motivated guests will have the ability to enroll in Martian courses to become more comfortable as well as actually understand the city and its host planet's fundamental information. Additionally, visitors can take part in research experiments as sort of luxurious guinea pigs living out a very unique few years of life. Of course, if particularly motivated, those only considered as visitors may be persuaded by the challenge they see from the workers of City NPC and choose to join as an intended full-time inhabitant. Regardless, especially early on, these guests will pay a lofty price to attend these holidays, mainly due to the intrinsic value of transporting and caring for people to and on Mars, but also due to the high-class experience and adventure meant to take place. Specifically, a stay of four total Earth years would cost on average $5 million and if it takes even at the least 100 years to finalize the one million populated city-state, we can expect a minimum of 250,000 total guests in this period which would bring in $1.25 trillion.

**Resource**

While it is definitely certain that Mars has an abundance of precious metals to exhaustingly mine, it is important to identify a physical resource to sell that is a byproduct of a constant production method used in City NPC. This resource for City NPC will be the heavy isotope of hydrogen called deuterium. Deuterium is a very valuable source of fuel for nuclear reactors, while also having the capability to be used in the production of heavy water which in turn can be used as fuel for fission reactors. It is deemed valuable because of the incredible advantage in regard to abundance it has over the Earth. On Mars, deuterium occurs 833 times out of every million hydrogen atoms, while it only comprises 166 out of every million hydrogen atoms on Earth giving Mars a near five-to-one advantage. Additionally, considering that we will have majority of our power used in water electrolysis to run our various life support needs, we will be able to produce deuterium at virtually no extra cost. Deuterium is a by-product of electrolysis which would produce around one kilogram of deuterium for every six tons of water. As opposed to the colony, City NPC will make great use and profit exporting this resource considering the heavily populated city's expansive use of electrolysis. For a city of one million, six million metric tons of water will be produced in one Earth year, allowing for one million kilograms of deuterium to be produced. Considering that pure deuterium is valued at approximately $10,000 per kilogram, this leaves us with a profit of around $9.8 billion after export costs. In the one hundred or so years it takes to finalize a city of one million people, 50 million kilograms of deuterium will be produced totaling to a profit of nearly $500 billion dollars after taxes using minimally exhaustive methods.

**Research and Media**

While research and media may be nicely suited for the initial colony of Mars, it will still serve as a prevalent business for the city. Billions of dollars will be invested from Earthling companies and institutions to hire thousands of Martians to garner highly sought-after data in a variety of fields. Just to imagine the race for Earthling companies even hundreds of years after humans came to Mars to inherit cutting edge technology unearthed by Martians justifies the potential here.

It is no strange act for television companies to capitalize off of extraordinary events and there will be one no bigger than the progression of a Martian city. One can only imagine that media exports would include City NPC's own leading television chef - amazing audiences by their unique Martian meals, or even a 'do-it-yourself' channel highlighting the limit of self-sustaining home technology that Earthlings could possibly make habit of. It would be without surprise if Mars found itself its own reality shows spawned by celebrity residents of the city's luxurious skyscrapers, or even its own Martian live streamers logging their day-to-day lives bringing in tons of sponsors. Quite frankly, these *Martian Media* ideas would be endless.

**Asteroid Mining**

The most audacious Martian industry will be that of asteroid mining. Mars is significantly closer to the asteroid belt than Earth, as the phenomena is slightly beyond Mars' orbit, with even those classified as Near-Earth-Objects actually orbiting closer to Mars, instead. The present-day issue isn't with the potential profitability of the industry, but logically due to the difficulty in collecting the asteroids and therefore their precious resources. There have been sample-return missions including Japanese unmanned spacecraft Hayabusa2 that have succeeded with firing a tantalum "bullet" at the asteroid called Ryugu and collected a piece. Additionally, the NASA initiative, *Asteroid Redirect Mission (ARM)*, that was planned for December 2021 but canceled due to budget cuts, involved two forms of resource capture: asteroid retrieval and asteroid redirection. The former involves sending a spacecraft to grip and retrieve fractional sizes of the asteroid, while the latter involves impacting, trapping, or even utilizing a concept known as a gravity tractor which makes use of a spacecraft's gravitational field to transmit the required impulse needed to redirect the asteroid. Considering the progression of technology for retrieval missions is more prevalent today on Earth, City NPC will assume to only utilize its technique while redirection will be reserved for a Martian country hundreds of years down the road.[9]

Setting up a rudimentary colony on Deimos will allow for the profitability of the asteroid mining business. With essentially no gravity to worry about, the Martian moon will provide great efficiency towards the propulsions and lift-offs needed for mining. Deimos is also thought to possess water ice at depth, relevant for refueling of liquid hydrogen intensive spacecraft. Considering the remaining surface area left available on Mars and the limited space on Deimos, missions will look to retrieve asteroids to desolate Martian areas, where they will then be transported to a Martian excavation site to mine for exportable goods.

Combining specified resource collection, research and media, tourism, and eventual asteroid mining, City NPC will be able to turn a profit quickly before the population passes one million. Table 1 details the economic progression over the period necessary to welcome one million people to City NPC.

Table 1: Summary of the financial consideration for City NPC.

| In Billion USD | Expense | | | Return | | | Yield |
|---|---|---|---|---|---|---|---|
| Year | Salary | Import | Investment | Asteroid Mining/Resource | Tourism | Research and Media | |
| 2065 | 100 | 50 | 5 | 50 | 0 | 200 | 95 |
| 2075 | 100 | 50 | 5 | 50 | 0 | 100 | -5 |
| 2085 | 150 | 25 | 5 | 50 | 0 | 100 | -30 |
| 2095 | 200 | 25 | 5 | 50 | 0 | 100 | -80 |
| 2105 | 250 | 15 | 3 | 50 | 0 | 100 | -118 |
| 2115 | 300 | <1/2 | 3 | 100 | 350 | 100 | 197 |
| 2125 | 310 | <1/2 | 0 | 150 | 150 | 100 | 140 |
| 2135 | 320 | <1/2 | 0 | 250 | 200 | 100 | 230 |
| 2145 | 330 | <1/2 | 0 | 400 | 300 | 100 | 470 |
| 2155 | 350 | <1/2 | 0 | 600 | 350 | 100 | 700 |
| Total | 2410 | ~200 | 26 | 1750 | 1350 | 1000 | **~1475** |

It is necessary to point out important periods detailed in the progression above. As stated before, the state is expected to issue full Martian wages for the first one hundred thousand people over 10 years; the eventual progression of the private sector will drastically lower this ratio. The year 2115 will begin to witness spikes from new and completely developed industries such as tourism and asteroid mining. After accumulating expected debt of around $200 billion, both Earthling investors and City NPC will see significant return on investment. Additionally, it is important to note the tourism projection in 2115 and the research and media projection in 2065. The latter seeks a global spike because it is expected that initial media coverage and advertising deals will be offered in abundance; beyond the first decade, media attention will decrease but gradually level off and research grants and tasks will positively persist providing a stable level of income. The former is reasonable considering the urge of the wealthy to travel to Mars especially considering the long development process needed which positively will build suspense and garner interest. Finally, this table meets expense characteristics outlined previously at around $2.5 trillion; the return is projected to gather around $4 trillion dollars, leaving a profitable yield of approximately $1.5 trillion for City NPC.

## Societal and Cultural

Colony NPC was constructed of values both societally and culturally similar to those on Earth simply due to the heavy concentration put towards immediate technical progress and success. Upon the finalization over decades of years of City NPC, purely Martian society and culture will flourish if the correct structures and values are put in place to allow for this ignition. However, it is important to remember to implement a socially engineered society that also allows for a city that organically grows culturally - to keep intact a cohesive group of people who intrinsically enjoy life on Mars.

### Professions

The full-time citizens of City NPC will be required to train and acquire expertise in at least one profession that creates the path to absolute self-sustainability. While it will be encouraged for future Martians to carve out unique work if it fits a certain niche or identity they possess, there will already be a plethora of industries and crafts available for all to enter. As opposed to a standard colony design, most workers in a city of this size will find themselves comfortable giving maximal effort to a sole profession, whereas colonists would be required to involve in secondary jobs to meet initial needs.

**Food:** Specifically, the crop production industry will require many workers. Those active in this area would be required to coordinate and oversee production activities such as harvesting, irrigation, planting, chemical application, and insurance of crop quality.

**Medical:** The importance of the breakthrough of advanced medicine on Earth can never be downplayed, as its results have led to a vastly healthier and prolonged society. Doctors and nurses will be present to treat common injuries and illness. Medical research will be extensively done on effects of Martian gravity on the human body. Therapists and psychologists will serve use to mentally ill or at least suffering members, with studies performed on the effects the radical challenge of planetary colonization has on the brain.

**Energy:** Maintaining the power grid on Mars will be vital to the enhancement of the city. Many engineers will inhabit the nuclear plants as well as field workers operating the surface level minimal solar grids with robotic assistance. Research will be done to maximize solar energy input, to perhaps one day convert the planet to a solely solar-powered existence.

**Educators:** Teaching on Mars will be held to a higher standard than Earth. As mentioned in the *education* section, Martian children will spend less time in school, but will be pushed at a rigorous enough pace as well as interact with challenges that are integral to success as Martians. Teachers on Mars will be required to have prior experience on the planet as a productive member to convey the knowledge necessary the young Martians will not take for granted, while not necessarily being required to have extensive teaching experience but more importantly a caring curriculum that shows they will exhaust all efforts to prepare the future.

**Scientists:** Scientists on Mars will be diversified to work on City NPC or Earth related tasks. Companies from Earth will ask botanists and chemists to research affects Martian soil has on terrestrial crops as well as studying what an unprecedented influx of carbon dioxide will do to accelerate production. Planetary scientists and geologists will study the composition of the Martian atmosphere, surface, and subsurface many degrees better than previous unmanned missions could do.

**Manufacturing:** Citizens working under the manufacturing industry will need to construct parts needed for housing and dome creation. Also, there will need to be specialists controlling and guiding 3D print construction, as well as CAD software engineers crafting the required designs.

**Additional:** Noting that a city of one million inhabitants would begin to allow for freedom and expression of trade, some niche careers will pop up. Restaurateurs, artists, fashion and interior designers and many others will all be welcomed to ignite their respective industries in City NPC.

**Martian Education**

Education in City NPC will consist of the technical and cultural necessities that allows for the survivability on Mars. While this schooling for young Martians should of course still retain the basic mathematics, language (perhaps a new one will arise), and history of Mars and Earth alike, the education should be integrated in the students' lives. There should be great emphasis on what brought them to this wonderful planet, why being a Martian is so important. The technical studies should be in touch with the environment around them, teaching these kids how and why what they learn will be useful for them to perhaps one day colonize a planet in another solar system, and not just learn about what a concept is. As on Earth, the need to educate youth properly to allow them to uphold values as well as knowledge that will let them use all of their potential to positively impact society is a must, so hopefully Mars will display that with great quality.

Martians will undergo schooling until they reach their early teens, where they will then take on remedial work or undertake apprenticeships. During this period, they will have a few years to cycle through the knowledge learned and after their relatively rudimentary day-work they will have time to apply this information to the world around, with the exceptionally bright ones creating the cutting edge technology at a rate unique to early day Mars. It is important to implement this progression, to not 'burn-out' the up-and-comers with distracting but necessary intensive work and instead allow for time to process and create.

The school-day will start early, but not be as prolonged as the present-day system. Only four-to-five hours is necessary with a direct education system, leaving down time for necessary physical activity and cultural experience to take place instead. There will not be a set school year, but instead only the natural progression of each student will determine the amount of time needed in their education journey. No higher-education institutions will be constructed, as the more higher-level tasks and opportunities will be available initially in primary system.

Specifically, Martian children will be taught the basics of language, mathematics, physics, and biology at a young age by their parents until they are enrolled in official schooling aged six. From there, students will begin with intermediately classed curriculum of those subjects, as well as introduction to basic computer programming and lab work. The next step would to undergo the advanced curriculum and begin to understand all intricacies of programming and research, until reaching a point where they can consider specializing slightly in a topic. These final years will see them take high-level courses in their preferred area of specialization, but still retain a schedule that includes sort-of capstone courses of the general topics that really focuses on integration to the technical society surrounding them.

**Calendar**

While it is important to allow social constructions like language to occur organically, and while the knowledge of the Earthling calendar and year is appropriate, implementing a Martian calendar will be integral in allowing the planet's citizens to form a unique identity.

The calendar of choice will be a simplified version of the Darian calendar first proposed by Thomas Grangale in 1985. The Martian year can be split into 24 total months, with the first 20 consisting of 28 sols and the remaining four consisting of 27 sols. Each week will consist of seven sols, with the necessary number of names destined to be determined by the initial inhabitants. An accommodation will need to be made for leap years, where six will occur every decade; the final month in each leap year will add a 28$^{th}$ sol, causing the year to be 669 Martian days long as opposed to a normal 668-sol year.[10, 11, 12]

**Principles of Martian Society**

It can be easy to forget that these brave inhabitants are not *robots* and will face physical, mental, and emotional queries. It is important to understand that perhaps the idealistic society is not realistic, and to factor in basic pillars citizens of City NPC will adhere by to allow for a community wound together by peace, resilience, and pride.

**Pillar 1 – Unity:** Success on Mars will live and die upon the community's motivation to come together and stay together. During difficulties it will be wise to not be afraid of relying on neighbors, friends, and coworkers: this cohesion will be necessary to overcome adversity. Likewise, it is imperative to remain humble and grateful during eras of great success; any development of greed or excessive grandiose will lead to separation and eventual collapse. Martians should have the mindset, perhaps not literally, but in the essence of metaphor: to live and die with and for each other.

**Pillar 2 – Well-being**: In order to extract high-level intrinsic motivation from an individual, physical; emotional; and psychological well-being needs to be positively present. To push the Martian citizen over the line of success or even progression, there needs to be enjoyment in the efforts to achieve these feats. Positivity and reassurance will come from encouragement from relationships, both intimate and professional, and the ability to accept eventual minor setbacks and adversity. To avoid the encourage of, especially, mental well-being will lead to great disruption throughout the city and eventually major setbacks and failures.

**Pillar 3 – Introspection and Self-accountability**: While Mars will for some time be known for its second-to-none unity, simultaneously its extraordinary individualism will be on display. The Martian should accept complete responsibility for their actions and be solely willing to make right and not wrong. On a daily basis, citizens should mull their desires and aspirations to keep up to date on their intended purpose and involvement. In order to remain consistently fit for duty in every sense, an individual cannot be pulled down by any regrets or mistakes weighing on them, but to search even deeper inside to understand why they're on this planet and why nothing will cause a loss of hope.

 **Pillar 4 – Expression:** The most encouraged trait to *express* in City NPC is that of expression. All citizens are pressed to follow through with their unique creations, even if doubt or comfort creeps in. Similar to unity, the difference between a mediocre yet successful and unequivocally thriving city is the ingenuity that lies within its inhabitants, so its oppression especially out of irrational fear will be frowned upon.

**Recreation and Leisure**

Both the quantity and quality of recreational activities available in a city are crucial to cultural and societal success. There should be no problem accommodating this necessity, especially considering the many possibilities inhabiting both tunnels and craters will bring.

Within the tunnels, there can be official and unofficial pod racing competition held on the parallel Hyperloop tracks. Whether it is just for fun or whether groups of inspired students actually design new pods and compete to see the winner, this unique event will be sure to bring in spectators from all over City NPC. Additionally, sanctioned rock-climbing events will be held along the 100-meter-high sides of a typical tunnel, with weak Martian gravity allowing extreme records.

In a crater is where team games will form, eventually leading to amateur leagues forming and played during festivities. Keeping in mind gravity, sports like basketball, baseball, and football will be an interesting watch, while sports like soccer and hockey will allow for more of an Earthling feel. To impose challenges for those somewhat more competitively individual minded, annual wrestling, boxing, and various martial arts tournaments will be held.

Finally, for those not fueled by intense sporting competition – spas, saunas and indoor and outdoor swimming pools will be installed allowing for a relaxing end to an evening; while upon darkness, films can be projected "miles" wide across the lower rims of the craters, creator a theater of thousands.

## An Aesthetically Pleasing City

While beauty is in the eye of the holder and every Martian could marvel at even a barren landscape spanning dozens of kilometers simply because of the opportunity it all possesses, effort will definitely be dedicated by the inhabitants of City NPC to make it an attractive and enjoyable place to live. Every Martian should be able to turn to an artifact or scene they find pleasant to comfort them in difficult times or even to find reassuring hope that the progress is paying off, at least visually. Specifically, Martians will possess personalized aesthetics to look upon, natural beauty from the landscape as a result of the site's location, and 'Martian-made' beauty spread throughout the city.

**Personalized**

Every Martian will be encouraged to bring along artifacts from home that offer hope or comfort, even if it is a favorite pair of underwear or t-shirt. Upmost importance will be put on enabling comfortability for the inhabitants of City NPC; it is obvious to understand that a comfortable and happy person usually relates to a productive and competitive one, but lying deeper is the intrinsic notion that a person submerged in pleasantries will reciprocally occupy their untouched surroundings with the same force. This can be related to a Martian community's drive to achieve terraformation: a city enhanced by its beauty and comfort they desire will cause them to accelerate efforts to expand these qualities.

Additional methods to achieve personal aesthetic is to determine the most sought-after domestic characteristics one can ask for. This will be different for each individual, but nonetheless we can detail typical desires; whether it is raising a favorite pet or living with or near close friends, painting murals upon the exterior of your level of the house or growing your own garden – City NPC will exceed any previously constructed hospitality principles.

**Natural**

Because of the naturally flat landscape accompanying the Arcadia Planitia region, bordering regions that surround will be available to marvel at. Located south west to the city with a safe distance between, but still close enough to gaze upon, is the large shield volcano Olympus Mons. Known as the largest volcano in the solar system will be enough to stun the Martians upon every glance with beauty, it can serve an even deeper purpose. Leaders in City NPC will encourage fellow inhabitants to few the landmark symbolically as its journey will resemble their own. They will be reminded that like themselves, the volcano began subsurface merely as flowing magma, in fact in a lava tube similar to their own, constantly erupting and expanding its dominance over much time eventually becoming something to marvel at. They will be reminded, perhaps in tongue-in-cheek fashion but nonetheless inspiring, that the volcano represents full self-sustainability: simply succeeding with what its environment offers.

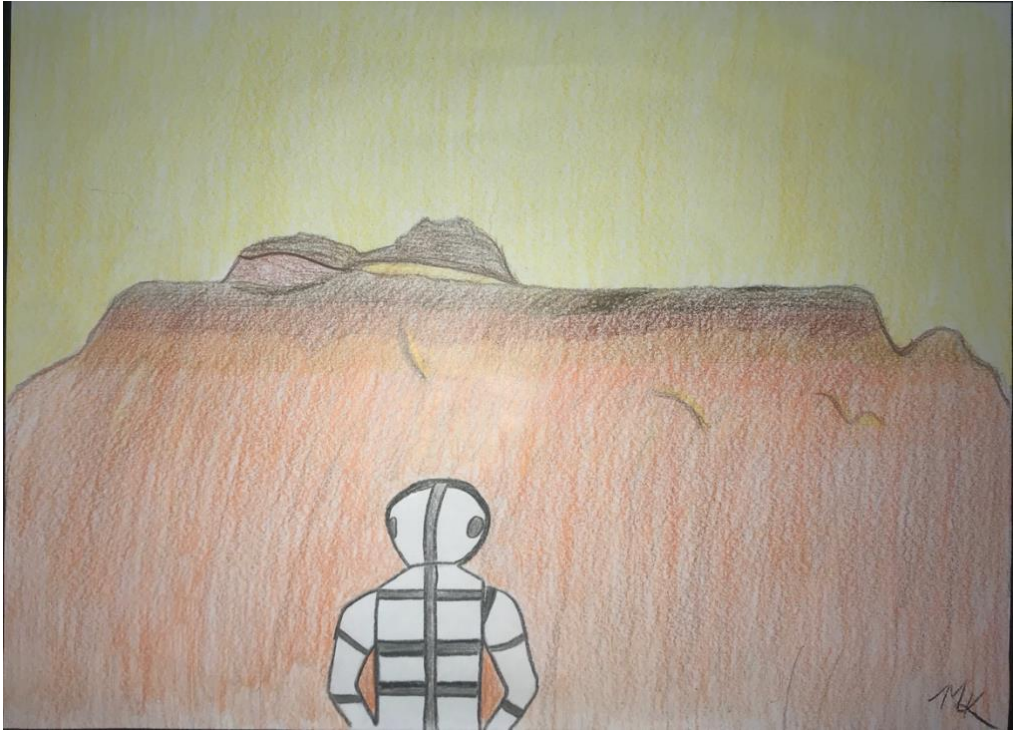
Figure 5: A Martian gazing out towards Olympus Mons.

**Martian-made**

The transition from rudimentary colony to growing city gives way to an influx of freedom and creativity from its citizens. While it is still very important to be as self-sufficient and financially savvy as possible, great importance lies in the additional steps to beautify the city itself.

The two areas of living both offer a plethora of ways to extraordinarily adorn the environment. Martians inhabiting the tunnels will have dozens of kilometers of lava tube wall to daub with fantastic displays of all imaginable disciplines of art. Perhaps, contests will be conducted to determine the most creative piece, surely increasing competitive edge and fulfilling the winners with much joy. Of course, those found unlucky will still be able to marvel at their own and many other beautiful and thoughtful creations producing certain joy. These inhabitants will eventually find themselves on the verge of producing self-sustainable Martian trees and native plants, possibly attempting to grow some upon the roof of their house that may be on its way in reaching the roof of the cave, some 50 meters up – a hilarious yet amazing sight to witness. Regardless, these underground-inhabiting citizens may just find themselves forgetting they are exactly that.

Realistically, the inhabitants of the Martian craters will have the upper hand when it comes to aesthetic. The circular area and subsequent dome already remind them of the landscape found when looking upwards at the Earth's sky. Undoubtedly, though, are the unique possibilities that present each community; whether it is the previously mentioned large-scale makeshift theaters or the eventual implementation of parks and wildlife zones.

Utilizing the glass panels occupying the geodesic dome, to derive maximum aesthetic potential, would to allow each and every citizen of City NPC to paint a symbolic gesture meaningful to them before construction. For example, all 200,000 occupants per sub city can paint their own distinct, but star-like symbol onto a glass panel. Upon complete construction, each individual would be astonished looking at the diversity of the mural. Ideally, each inhabitant of City NPC will look upon that symbol for personal guidance during difficult times, similar to lost adventurers relying on the North Star for directional guidance.

## Conclusion

Establishing a Martian city-state full of the fundamental precision to not only survive but succeed presents a multitude of challenges. Careful consideration will need to go into the habitable planning of the city, to ensure a relatively high quality. This consideration also presents the inhabitants with an advantage in the production of various in-situ processes. Implementing reasonable, yet profitable financial and economic systems will ensure the opportunity for the city to garner the funds needed to survive. Making sure the entire society as well as its sub-communities find pleasure in their cultural makeup will enable a cohesive group fighting together as opposed against is vital. Critically caring for all these aspects of City NPC will be crucial in its fight for self-sustainability and more importantly its forever lasting impression on Mars.

## Acknowledgment

The authors would like to acknowledge and give thanks to The Mars Society for the opportunity to participate in the Mars City State Contest "Design the First 1-Million Person City on Mars."

## References

1. Mosher, David. (2019). SpaceX is eyeing these 9 places on Mars for landing its first Starship rocket missions. Accessed February 13, 2020, https://www.businessinsider.com/spacex-starship-mars-landing-sites-map-hirise-2019-9.
2. USGA Maps. (2015). MOLA map showing boundaries for part of Arcadia Planitia and other regions. Accessed March 27, 2020, http://planetarynames.wr.usgs.gov/images/mola_regional_boundaries.pdf.
3. Sloat, Sarah. (2017). Martian Colonists Could Live in Lava Tubes Beneath the Surface. Accessed January 30, 2020, https://www.inverse.com/article/36777-mars-moon-human-colony-lava-tubes.
4. AI Spacefactory. (2019). Marsha AI Spacefactory's Mars Habitat. Accessed April 10, 2020, https://www.aispacefactory.com/marsha.
5. Taylor, G. J. (2009). Mars Crust: Made of Basalt. *Planetary Science Research Discoveries.* Accessed January 14, 2020, http://www.psrd.hawaii.edu/May09/Mars.Basaltic.Crust.html.
6. IMSR Technology. (2020). Integral Molten Salt Reactor: Safe, clean, low-cost, high-impact and resilient. Accessed June 10, 2020. https://www.terrestrialenergy.com/technology/.
7. Nield, David. (2019). Peas, Quinoa And 7 Other Crops Grown Successfully in Soil Equivalent to Moon And Mars. Accessed February 10, 2020, https://www.sciencealert.com/experiments-show-soil-from-the-moon-and-mars-could-support-crops.
8. HYBRIT. (2020). Description of clean steelmaking using hydrogen instead of coal. Accessed May 8, 2020.
9. Physics World. (2018). The Asteroid Trillionaires. Accessed March 15, 2020, https://physicsworld.com/a/the-asteroid-trillionaires/.
10. Smith, Arthur E. (1989). Mars: The Next Step, p. 7. Taylor & Francis. Accessed February 13, 2020.
11. Gangale, Thomas. (1986). Martian Standard Time. Journal of the British Interplanetary Society. Vol. 39, No. 6, p. 282-288. Accessed February 13, 2020.
12. Gangale, Thomas. (2006). The Architecture of Time, Part 2: The Darian System for Mars. Accessed February 14, 2020, https://saemobilus.sae.org/content/2006-01-2249.